\pdfoutput=1

\documentclass[11pt,twoside,a4paper,cmspaper,final,collab]{cms-tdr}
\def\svnVersion{454081P}\def\cmsCernNoTag{CERN-EP-2017-261}\def\cmsCernDate{\today}\def\cmsMessage{Published in the Journal of High Energy Physics as \href{http://dx.doi.org/10.1007/JHEP01(2018)045}{\doi{10.1007/JHEP01(2018)045.}}}

\begin{document}\cmsNoteHeader{HIN-16-021}

\hyphenation{had-ron-i-za-tion}
\hyphenation{cal-or-i-me-ter}
\hyphenation{de-vices}
\RCS$Revision: 454059 $
\RCS$HeadURL: svn+ssh://svn.cern.ch/reps/tdr2/papers/HIN-16-021/trunk/HIN-16-021.tex $
\RCS$Id: HIN-16-021.tex 454059 2018-04-04 19:42:27Z rbi $
\newlength\cmsFigWidth
\ifthenelse{\boolean{cms@external}}{\setlength\cmsFigWidth{0.85\columnwidth}}{\setlength\cmsFigWidth{0.4\textwidth}}
\ifthenelse{\boolean{cms@external}}{\providecommand{\cmsLeft}{top\xspace}}{\providecommand{\cmsLeft}{left\xspace}}
\ifthenelse{\boolean{cms@external}}{\providecommand{\cmsRight}{bottom\xspace}}{\providecommand{\cmsRight}{right\xspace}}

\newcommand {\pp} {pp\xspace}
\newcommand {\pA} {pA\xspace}
\newcommand {\NN} {AA\xspace}
\newcommand {\pPb} {pPb\xspace}
\newcommand {\sqrts} {\ensuremath{\sqrt{s}}\xspace}
\providecommand {\sqrtsNN} {\ensuremath{\sqrt{\smash[b]{s_{_{\mathrm{NN}}}}}}\xspace}
\newcommand {\Nch} {\ensuremath{N_{\text{ch}}}\xspace}
\newcommand {\etacm} {\ensuremath{\eta_{\text{cm}}}\xspace}
\newcommand {\etalab} {\ensuremath{\eta_{\text{lab}}}\xspace}
\newcommand {\dnchdeta} {\ensuremath{\rd{}N_{\text{ch}}/\rd\eta}\xspace}
\newcommand {\dnchdetalab} {\ensuremath{\rd{}N_{\text{ch}}/\rd\eta_{\text{lab}}}\xspace}
\newcommand {\dnchdetacm} {\ensuremath{\rd{}N_{\text{ch}}/\rd\eta_{\text{cm}}}\xspace}
\newcommand {\avednchdetacm} {\ensuremath{\langle\rd{}N_{\text{ch}}/\rd\eta_{\text{cm}}\rangle{\vert_{\abs{\eta_{\text{cm}}} < 0.5}}}\xspace}
\newcommand {\npart} {\ensuremath{N_{\text{part}}}\xspace}
\newcommand{\EPOS}{{\textsc{Epos}}\xspace}
\newcommand{\hijing}{{\textsc{Hijing}}\xspace}
\newcommand{\dpmjet}{{\textsc{Dpmjet}}-III\xspace}

\cmsNoteHeader{HIN-16-021}
\title{Pseudorapidity distributions of charged hadrons in proton-lead collisions at $\sqrtsNN = 5.02$ and 8.16\TeV}

\date{\today}

\abstract{
The pseudorapidity distributions of charged hadrons in proton-lead collisions
at nucleon-nucleon center-of-mass energies $\sqrtsNN = 5.02$ and 8.16\TeV are
presented. The measurements are based on data samples collected by the CMS
experiment at the LHC. The number of primary charged hadrons produced in
non-single-diffractive proton-lead collisions is determined in the
pseudorapidity range $\abs{\etalab} < 2.4$. The charged-hadron multiplicity
distributions are compared to the predictions from theoretical calculations and
Monte Carlo event generators. In the center-of-mass pseudorapidity range
$\abs{\etacm} < 0.5$, the average charged-hadron multiplicity densities
$\avednchdetacm$ are $17.31 \pm 0.01\stat\pm 0.59\syst$
and $20.10 \pm 0.01\stat\pm 0.85\syst$ at $\sqrtsNN
= 5.02$ and 8.16\TeV, respectively. The particle densities per participant
nucleon are compared to similar measurements in proton-proton, proton-nucleus,
and nucleus-nucleus collisions.
}

\hypersetup{%
pdfauthor={CMS Collaboration},%
pdftitle={Pseudorapidity distributions of charged hadrons in proton-lead collisions at sqrt(s[NN]) = 5.02 and 8.16 TeV},%
pdfsubject={CMS},%
pdfkeywords={CMS, physics, heavy-ion, proton-lead}}

\maketitle
\section{Introduction}

Studies of charged-hadron yields have long been a key tool for exploring
perturbative and nonperturbative quantum chromodynamics (QCD) phenomena in
high-energy particle and nuclear collisions~\cite{Busza:1975te}. Measurements
in proton-lead (\pPb) collisions can shed light on initial-state nuclear
effects in these interactions~\cite{Salgado:2011wc}. An example is the nuclear
modification of parton distribution functions (PDFs) that can be observed in
measurements of hadron~\cite{Khachatryan:2015xaa,Khachatryan:2016odn,
Khachatryan:2015uja,Adam:2016ich,Sirunyan:2017mzd} and
jet~\cite{Khachatryan:2015sva,Khachatryan:2016xdg,Chatrchyan:2014hqa}
production. Such measurements also provide reference data for understanding the
hot, dense medium produced in nucleus-nucleus (AA) collisions. At the CERN LHC
energies, measurements of proton-nucleus (\pA) collisions allow studies of the
nuclear gluon distributions and parton shadowing effects at very small values
($10^{-4}$--$10^{-6}$) of the Bjorken $x$ variable~\cite{Salgado:2011wc,
Dumitru:2011wq}. This provides a crucial test of current theoretical approaches
for high-energy QCD~\cite{Dumitru:2011wq,Xu:2012au,Albacete:2012xq}, and yields
important constraints on phenomenological models and event
generators~\cite{Wang:1991hta,Gyulassy:1994ew,Roesler:2000he,Pierog:2013ria}.

The number of primary charged hadrons, \Nch, is commonly characterized by its
pseudorapidity density, \dnchdeta. The pseudorapidity, $\eta$, is defined as
$-\ln[\tan{\theta/2}]$, where $\theta$ is the polar angle of the particle with
respect to the beam axis. The center-of-mass energy dependence of \dnchdeta 
constrains the theoretical modeling of particle production arising from hard and soft 
QCD processes in high-energy hadronic interactions.  In the
presence of the quark-gluon plasma (QGP), the hot medium produced in AA
collisions, modifications of hadron production have been observed. Studying the
energy dependence of the pseudorapidity density in different colliding systems
(proton-proton (pp), pA, AA), for both total inelastic and
non-single-diffractive (NSD)~\cite{Halliwell:1977tb,Kittel:2005fu,
Khachatryan:2010nk} collision processes, improves our understanding of these
modifications in the AA case by identifying nuclear effects present in the
initial state. Monte Carlo (MC) event generators, which reproduce the main
characteristics of experimental results from hadronic collisions at lower
energies, can provide predictions for the energy dependence of hadron
production using different implementations of QCD
effects~\cite{d'Enterria:2011kw}.

In this paper, measurements of \dnchdetalab (where the pseudorapidity is
measured in the laboratory frame) in the range $\abs{\etalab} < 2.4$ are
reported for NSD events in \pPb collisions delivered by the LHC in 2016 at
$\sqrtsNN = 5.02$ and 8.16\TeV. Following earlier analyses in pp collisions at
$\sqrts = 0.9$--13\TeV~\cite{Khachatryan:2010xs,Khachatryan:2010us,
Chatrchyan:2014qka,Khachatryan:2015jna} and in lead-lead collisions at $\sqrtsNN
= 2.76$\TeV~\cite{Chatrchyan:2011pb}, \Nch is restricted to ``primary" charged
hadrons, defined to include prompt hadrons as well as decay products of all
particles with proper decay length $c\tau < 1\unit{cm}$, where $\tau$ is the
proper lifetime of the particle and $c$ is the velocity of light in vacuum.
Contributions from prompt leptons and decay products of longer-lived particles
and secondary interactions are excluded. For $\sqrtsNN = 5.02\,(8.16)$\TeV, the
beam energies per nucleon were 4 (6.5)\TeV and 1.58 (2.56)\TeV for the proton
and lead nucleus, respectively. Because the beam energies were asymmetric and
the proton was going in the positive \etalab direction, massless particles
emitted at midrapidity in the nucleon-nucleon center-of-mass, $\etacm = 0$,
will be detected at $\etalab = 0.465$. Results are compared to predictions from
the KLN model~\cite{Dumitru:2011wq}, as well as the \EPOS LHC
(v3400)~\cite{Werner:2005jf,Pierog:2013ria}, \hijing~\cite{Wang:1991hta}
(versions 1.3~\cite{Gyulassy:1994ew} and 2.1~\cite{Xu:2012au}), and
\dpmjet~\cite{Roesler:2000he} MC event generators. The \sqrtsNN dependence of
\dnchdetacm in the region $\etacm \approx 0$ is also presented.

\section{The CMS detector}

The central feature of the CMS apparatus is a superconducting solenoid of
6\unit{m} internal diameter. Within the solenoid volume are a silicon pixel and
strip tracker, a lead tungstate crystal electromagnetic calorimeter, and
a brass and scintillator hadron calorimeter, each composed of a barrel and two
endcap sections. The silicon tracker measures charged particles within the
range $\abs{\etalab} < 2.5$. It consists of 1440 silicon pixel detector
modules. The barrel region of the pixel detector consists of three layers,
which are very close to the beam line. They are located at average radii of
4.3, 7.2, and 11.0\unit{cm}, and provide excellent position resolution with
their $150{\times}100\mum$ pixels. The forward hadron (HF) calorimeter uses
steel as an absorber and quartz fibers as the sensitive material. It consists
of two halves, each located 11.2\unit{m} from the interaction region, and
together they provide coverage in the range $3.0 < \abs{\etalab} < 5.2$. The
beam pickup for timing (BPTX) devices were used to trigger the detector
readout. They are located around the beam pipe at a distance of 175\unit{m} on
either side of the interaction point (IP) and are designed to provide precise
information on the LHC bunch structure and the timing of the incoming beams.
Muons are detected in gas-ionization chambers embedded in the steel flux-return
yoke outside the solenoid. A more detailed description of the CMS detector,
together with a definition of the coordinate system used and the relevant
kinematic variables, can be found in Ref.~\cite{Chatrchyan:2008aa}.

\section{Event selection}

The data used in this analysis were taken with the beam configuration in which
the proton beam traveled in the negative pseudorapidity direction, and selected
to contain collision events recorded during low-intensity beam configurations,
with 0.3--0.6\% proton-lead interaction probability per bunch crossing. The
collision events are selected online by requiring a coincidence of signals from
both BPTX devices, indicating the presence of both proton and lead ion bunches
crossing the IP, and at least one energy deposit above the readout threshold of
3\GeV on either side of the HF. The offline selection of NSD events is
accomplished by requiring that at least one energy deposit greater than 3\GeV
is found on each of the two sides of the HF and at least one reconstructed
interaction vertex is found. A study of noncolliding bunches shows that these
requirements are also sufficient to reject all backgrounds not originating from
\pPb collisions. The probability to select events in the presence of a single
(noncolliding) beam is found to be around $2 \times 10^{-5}$ per bunch
crossing, to be compared to the average number of collisions per bunch
crossing of $4.5 \times 10^{-3}$. Consequently, the contribution of background
events from beam, beam halo, and cosmic ray sources to the observed yields is
negligible. The total number of \pPb collision events passing the selection
criteria is approximately 420 thousand and 3 million at $\sqrtsNN = 5.02$ and
8.16\TeV, respectively.

The corrections from the detector-level offline event selection to the
hadron-level event definition are derived from MC simulations with the \EPOS
generator. The MC simulations are produced with the same vertex distribution
along the interaction region as observed in data. The detector response is
simulated with \GEANTfour~\cite{Agostinelli:2002hh} and processed through the
same event reconstruction chain as the collision data.

\section{Data analysis}
\label{sec:analysis}

In the presence of a magnetic field, charged particles follow curved
trajectories, perturbed mostly by multiple Coulomb scattering. The
reconstructed pixel clusters (or ``hits") alone are sufficient to reconstruct
vertices and tracks with high precision and purity. The analysis technique is
based on tracklets, pairs of hits from two different layers, and relies on the
fact that for a primary charged hadron, the differences in pseudorapidity
($\Delta\eta$) and azimuthal angle ($\Delta\phi$) between the two hits are
small. This method is sensitive to charged hadrons with transverse momenta
\pt as low as 40\MeVc.

The primary vertex reconstruction is based on pixel hits in the first two
layers of the detector, as in Ref.~\cite{Chatrchyan:2011pb}. In the first step,
a hit from the first layer is selected and a matching hit from the second layer
is sought. If the $\abs{\Delta\phi}$ of the hits is smaller than 0.05
(optimized to maximize the vertex reconstruction efficiency), the $z$ positions
of the hits (with the $z$ axis defined to be parallel to the beam axis) are
extrapolated linearly and projected onto the beam axis. This procedure is
repeated for every hit in the first layer, and the projected $z$ positions are
saved as vertex candidates. The primary vertex is determined in a second step.
If the magnitude of the difference between the $z$ positions of any two vertex
candidates is smaller than 0.12\unit{cm}, they are combined into a vertex
cluster. The vertex cluster with the highest number of associated vertex
candidates is selected as the primary vertex, and the final vertex $z$
position, $z_v$, is given by the average $z$ position of the associated vertex
candidates. The typical resolution of $z_v$ is 0.02--0.04\unit{cm}, depending
on the number of pixel hits. The vertex reconstruction efficiency is found to
be high even for low-multiplicity events with few pixel hits, with around 90
(100)\% efficiency for events with 4 (10) hits in the first layer.

\begin{figure}[thb]
\centering
\includegraphics[width=0.32\textwidth]{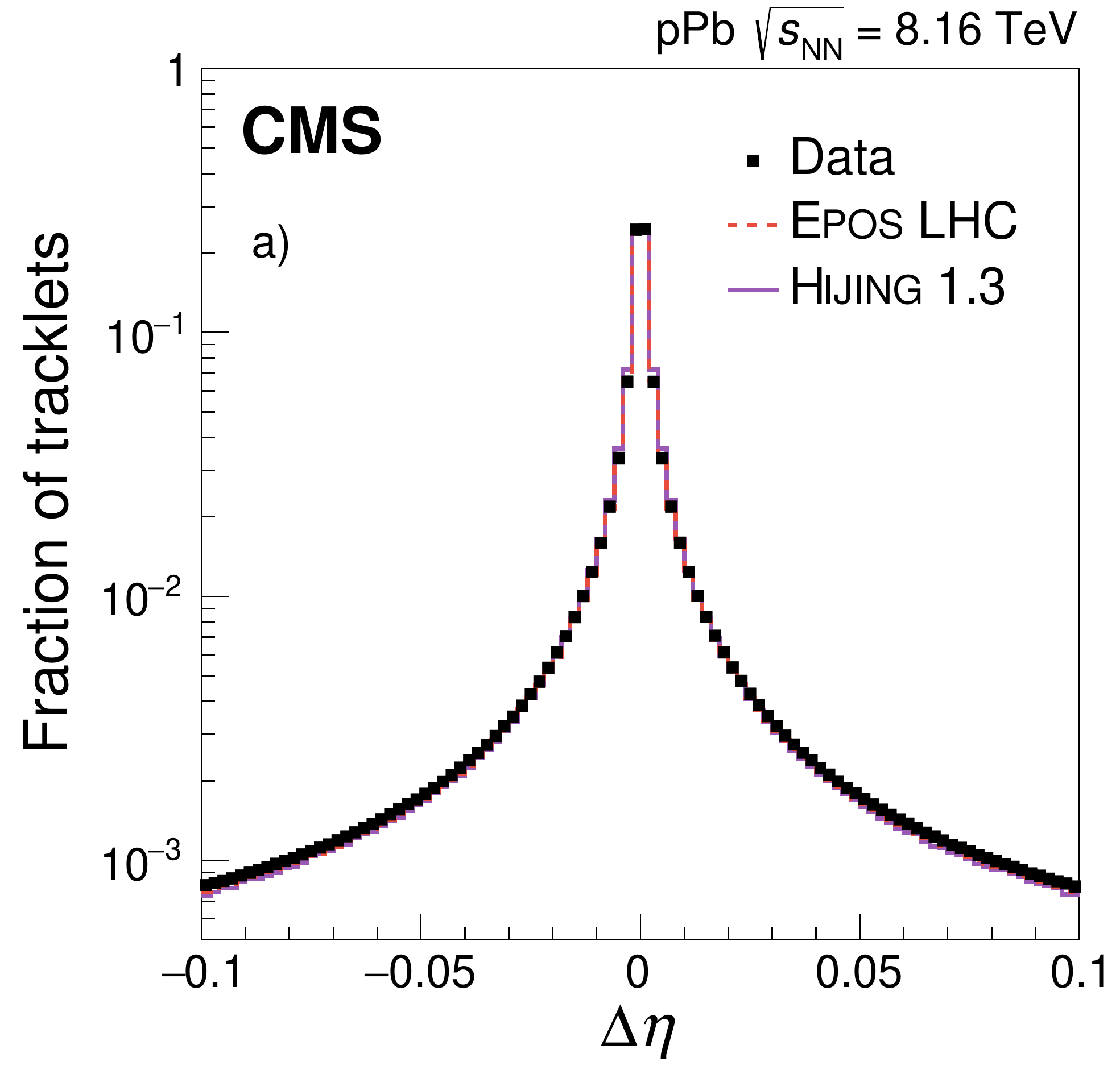}
\includegraphics[width=0.32\textwidth]{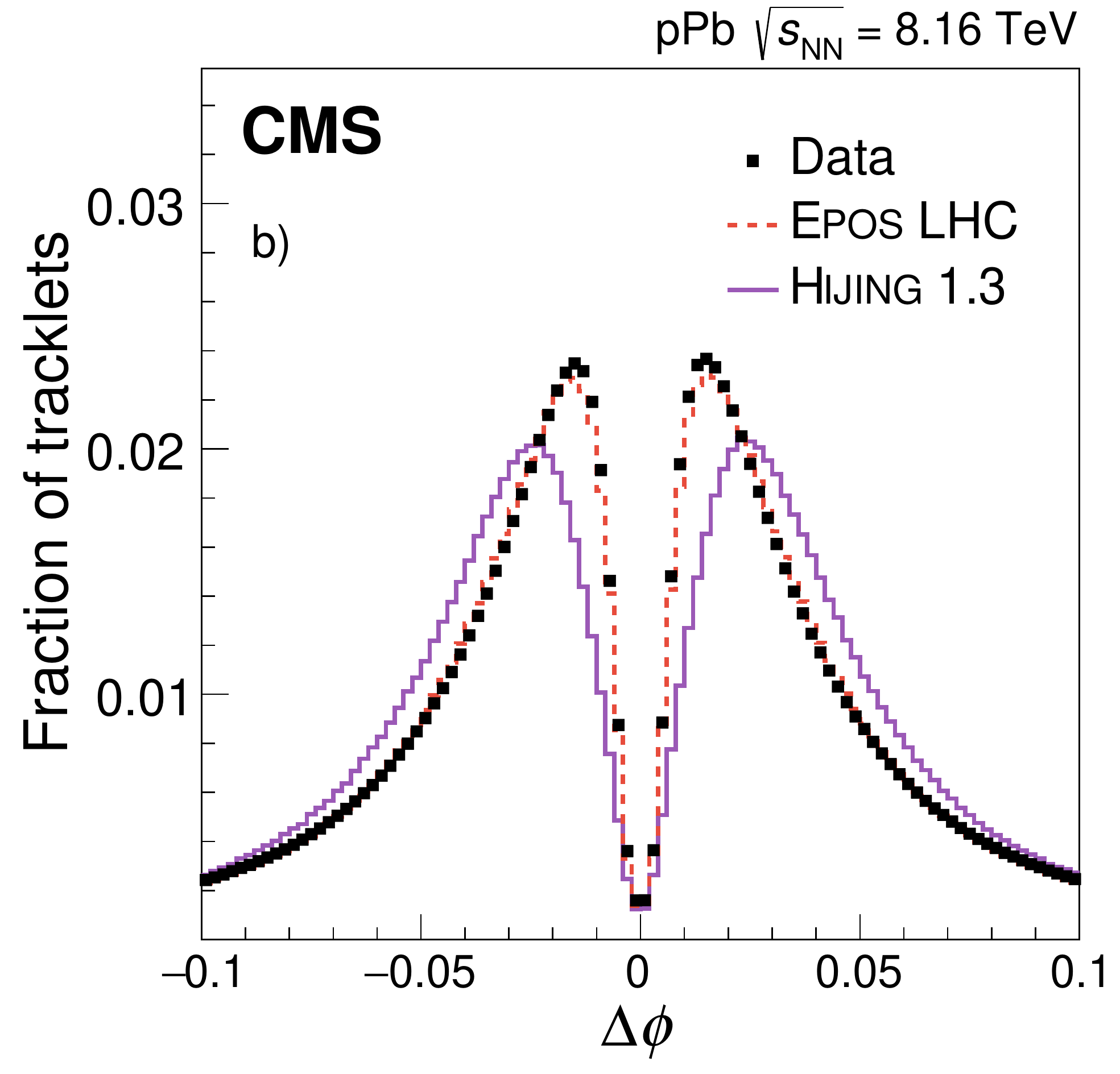}
\includegraphics[width=0.32\textwidth]{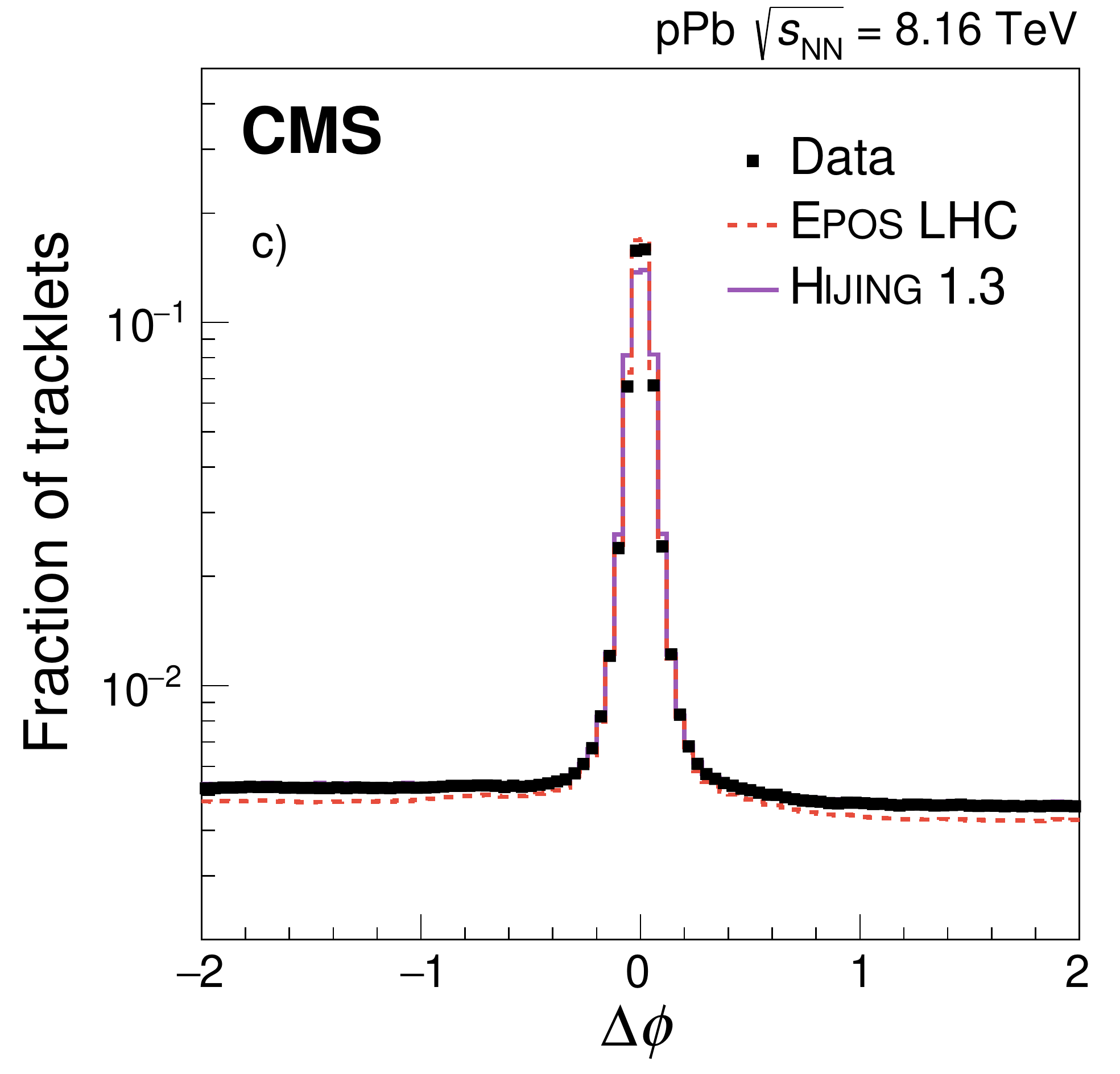}
\caption{The $\Delta\eta$ (a) and $\Delta\phi$ (b, c) distributions of hit
pairs for tracklets in \pPb collisions at 8.16\TeV (squares) and from MC
simulations with the \EPOS and \hijing~1.3 generators (solid lines). The
statistical uncertainties are smaller than the marker sizes for all
distributions shown. }
\label{fig:tracklet}
\end{figure}

The tracklet reconstruction follows a separate algorithm from the vertex
reconstruction. There is no requirement on the $\Delta\phi$ of the hits.
Instead, a hit on a given layer is paired with the hit on another layer which
is closest in $\eta$ (where $\eta$ is measured with respect to the primary
vertex) and these two hits form a tracklet. No hit can be used more than once.
No selection is applied on the hit quality or charge, such that the analysis
is rather insensitive to the accuracy of the simulation of pixel cluster
charge. Three different types of tracklets can be reconstructed, corresponding
to different combinations of the three pixel detector layers: 1+2, 1+3, and
2+3. The reconstruction efficiency, acceptance, fraction of background hits,
and sensitivity to particle \pt is different for each type of tracklet. This
serves as a consistency check for the analysis, and reduces systematic biases
in the measurement.

Figures~\ref{fig:tracklet}(a) and (b) show the $\Delta\eta$ and $\Delta\phi$
distributions of reconstructed hit pairs for tracklets in data and simulation.
To suppress the combinatorial background, while still including most particles
in the analysis, only tracklets with $\abs{\Delta\eta} < 0.1$ are considered
``signal''. In this kinematic region, there is good agreement between data and
simulations with the \EPOS generator, indicating that the $\pt$ distributions
of both hard and soft particles in data are described well by this MC
generator. The \hijing generator, used in this analysis for systematic studies,
gives a poorer description of the distributions, especially for $\Delta\phi$
. Tracklets corresponding to charged hadrons that originate from the primary
vertex have small but nonzero $\Delta\phi$ due to the magnetic field in the
detector, while background tracklets from uncorrelated pixel hits form
a roughly flat $\Delta\phi$ spectrum over the entire $\Delta\phi$ range, as
shown in Fig.~\ref{fig:tracklet}(c), where the abscissa is extended to
$\abs{\Delta\phi} < 2$. Hence, a sideband region defined by $1
< \abs{\Delta\phi} < 2$ is used to estimate the background fraction, which is
then subtracted from the signal region $(\abs{\Delta\phi} < 1)$ to obtain the
uncorrected $\dnchdetalab$~\cite{Chatrchyan:2011pb}. The background estimation
and subtraction is performed as a function of \etalab, $z_v$, and tracklet
multiplicity. Typical values of the estimated background fraction in the signal
region in data increase with $\abs{\etalab}$ from 10 -- 25\%. The \etalab range
is restricted to $\abs{\etalab} < 2.4$ to avoid a large acceptance correction.

The final results need to be corrected for contributions from decaying
particles with $c\tau > 1\unit{cm}$, particles created in secondary
interactions, and prompt leptons. The contribution of these particles to
\dnchdetalab is removed using a correction factor found using MC simulations.
In addition, corrections are needed to account for the selection, efficiency,
and acceptance of reconstructed tracklets, as well as trigger and vertexing
efficiencies. The acceptance factor includes the extrapolation down to $\pt
= 0\GeVc$. Correction factors (with a typical total of ${<}15\%$) are derived
using the \EPOS event generator as a reference and are calculated as a function
of \etalab, $z_v$, and tracklet multiplicity, as was done in
Ref.~\cite{Chatrchyan:2011pb}.

To account for the differences between data and MC in the pixel detector
geometry and its alignment conditions, an additional correction is applied as
a function of \etalab and $z_v$. This correction is obtained by taking the
ratio between data and simulation of the geometrical distribution of tracklets
in (\etalab, $z_v$) intervals. The size of this correction ranges from 0 to
5\%, where the largest correction factors are associated with the presence of
inactive tracker modules.

\subsection{Systematic uncertainties}

The systematic uncertainties in the final results arise from several sources:
detector misalignment, pixel hit reconstruction inefficiency, pixel cluster
splitting, background modeling, selection of signal and sideband regions,
parametrization of the correction factors, and the NSD event selection. For
each source of uncertainty, that part of the analysis procedure is varied
independently and the change is propagated to the final results. The individual
contributions are then summed in quadrature to give the total systematic
uncertainty.

To estimate the uncertainty from detector misalignment, each pixel hit is
offset by a small distance corresponding to the uncertainty in the alignment of
the pixel detectors. The effects of pixel hit reconstruction inefficiency are
studied by randomly excluding 0.5\% of the pixel hits from the analysis. The
0.5\% inefficiency value is determined by studying tracklets reconstructed from
pixel hits in layers~1 and 3, and taking the double ratio in data and
simulation of the fraction of tracklets that have no corresponding hit in
layer~2. Pixel cluster splitting refers to the situation where the charge
deposit in the pixel detector from a single charged particle is reconstructed
as two separate pixel clusters. Its effect on the measurement is estimated by
randomly splitting pixel clusters with a probability of 1.2\%, as determined by
previous studies~\cite{Khachatryan:2010xs}. The contributions from the above
three sources are all below 1\%.

The remaining uncertainties are associated with the MC correction factors.
Additional pixel hits, randomly sampled from the hit distributions in data, are
added such that the $\Delta\phi$ sidebands match between data and MC. The
percentage of additional pixel hits needed is less than 5\%. The variations
observed compared to the nominal results are around 1.5--2.5\%. The signal and
sideband regions are also varied to $\abs{\Delta\phi} < 1.5$ and $1.5
< \abs{\Delta\phi} < 3.0$, respectively. A variation of 0.6--1.5\% is found as
compared to the nominal setting, which is propagated as a systematic
uncertainty. Different multiplicity variables are used to parametrize the
correction factors, in addition to the background-subtracted tracklets variable
used for the nominal results: number of tracklets (before background
subtraction), number of pixel hits in the first pixel layer used (layer~1 for
tracklet type 1+2 and 1+3, and layer~2 for tracklet type 2+3). The maximum
deviation in each $\etalab$ interval, 1.5--2.5\%, is quoted as an uncertainty.
An uncertainty is assigned for the selection of NSD events. The fraction of the
single-diffractive events removed by the event selection, as determined from
the \EPOS generator, is 16\% when the tracklet multiplicity in the event is
less than 10, and falls quickly to 0\% with increasing tracklet multiplicity.
This fraction is varied from 0\% to twice the nominal value, and the maximum
deviation from the final results, 1.2\%, is quoted as the uncertainty.
A summary of the systematic uncertainties for the measurements at 5.02 and
8.16\TeV is shown in Table~\ref{table:systematics}.

\begin{table}[hbtp]
\begin{center}
\caption{Summary of the systematic uncertainties from various sources, for \pPb
collisions at 5.02 and 8.16\TeV. The range of values indicates the minimum and
maximum uncertainties across the \etalab range. }
\label{table:systematics}
\begin{tabular}{p{0.12\textwidth} p{0.45\textwidth} c c}
\hline
& \multirow{2}{*}{Source}                  & \multicolumn{2}{c}{Uncertainty [\%]}   \\
&                                          &   5.02\TeV         &   8.16\TeV        \\
\hline \hline
\multirow{3}{0.12\textwidth}{Data and simulation}
& {Detector misalignment}                  & {0.2 -- 1.0}       & {0.2 -- 1.0}      \\
& {Pixel hit reconstruction inefficiency}  & {1.0}              & {1.0}             \\
& {Pixel cluster splitting}                & {0.3 -- 0.8}       & {0.3 -- 0.6}      \\
\hline
\multirow{4}{0.12\textwidth}{MC corrections}
& {Background modeling}                    & {1.3 -- 3.2}       & {1.5 -- 2.5}      \\
& {Signal and sideband region selection}   & {0.5 -- 1.5}       & {0.6 -- 1.5}      \\
& {Choice of parametrization variable}     & {1.6 -- 2.5}       & {1.5 -- 3.5}      \\
& {NSD selection}                          & {1.2}              & {1.2}             \\
\hline \hline
& {Total uncertainty}                      & {3.0 -- 4.3}       & {3.7 -- 4.6}      \\
\hline
\end{tabular}
\end{center}
\end{table}

\section{Results}

\begin{figure}[t!bhp]
\centering
\includegraphics[width=0.55\textwidth]{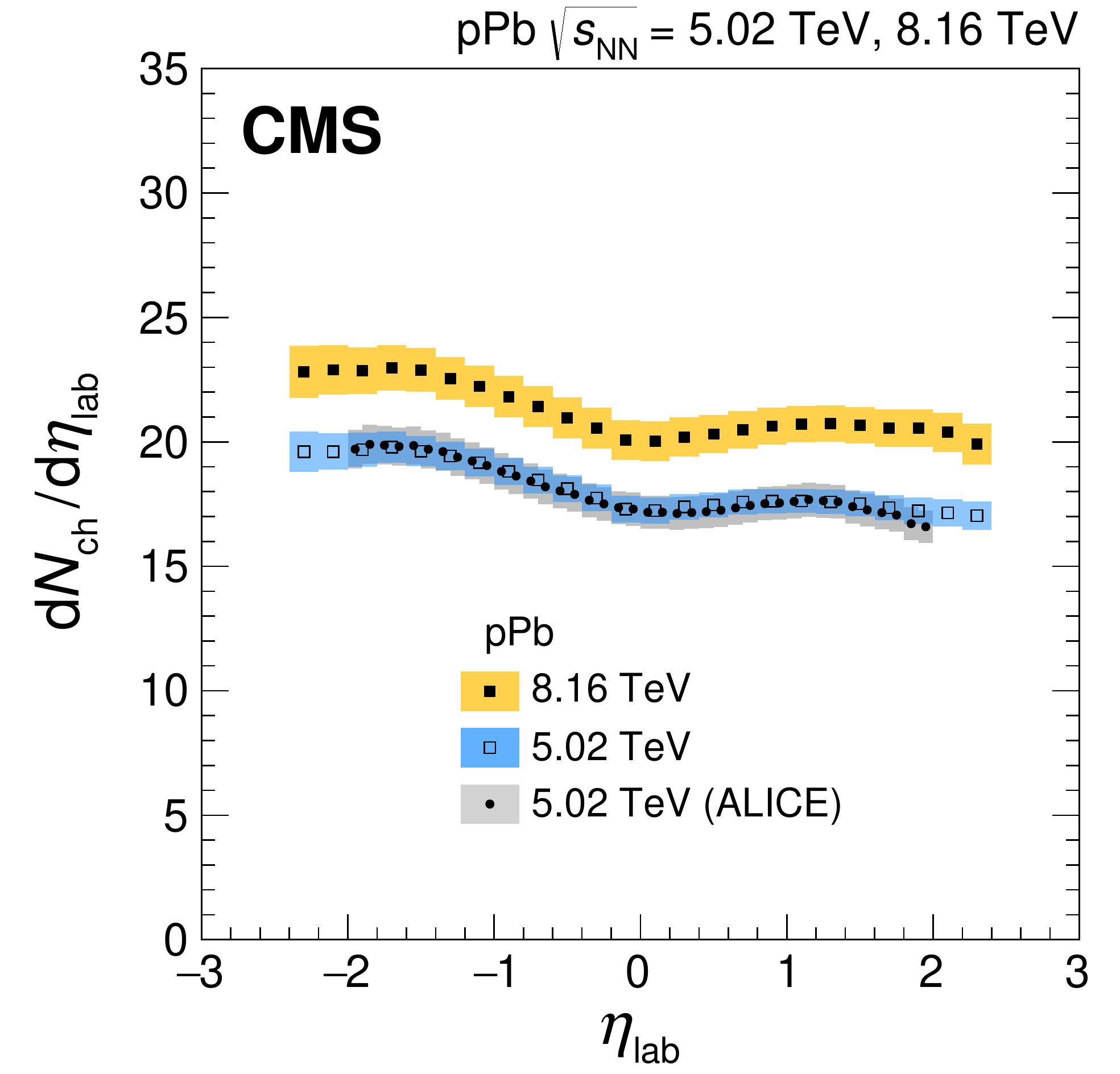}
\caption{Distributions of the pseudorapidity density of charged hadrons in the
region $\abs{\etalab} < 2.4$ in NSD \pPb collisions at $\sqrtsNN = 5.02$ (open
squares) and 8.16\TeV (full squares). The measurement at 5.02\TeV by the ALICE
Collaboration~\cite{ALICE:2012xs} is shown as filled circles. The shaded boxes
indicate the systematic uncertainties which, in the case of the CMS data, are
correlated between the two beam energies. The proton beam goes in the positive
\etalab direction. }
\label{fig:dndeta_final}
\end{figure}

Pseudorapidity density distributions of charged hadrons in the region
$\abs{\etalab} < 2.4$ for NSD \pPb collisions are shown in
Fig.~\ref{fig:dndeta_final}. The distributions shown are the average of the
measured distributions from the three types of tracklets (1+2, 1+3, and 2+3),
which are consistent with each other within 3\%. A clear difference in the
particle densities between the lead ion ($\etalab < 0$) and the proton
($\etalab > 0$) beam directions is observed. The measured \dnchdetalab
distribution at 5.02\TeV agrees with the measurement by the ALICE
Collaboration~\cite{ALICE:2012xs}. The multiplicities at 8.16\TeV are
significantly higher than those at 5.02\TeV.

\begin{figure}[t!bhp]
\centering
\includegraphics[width=0.55\textwidth]{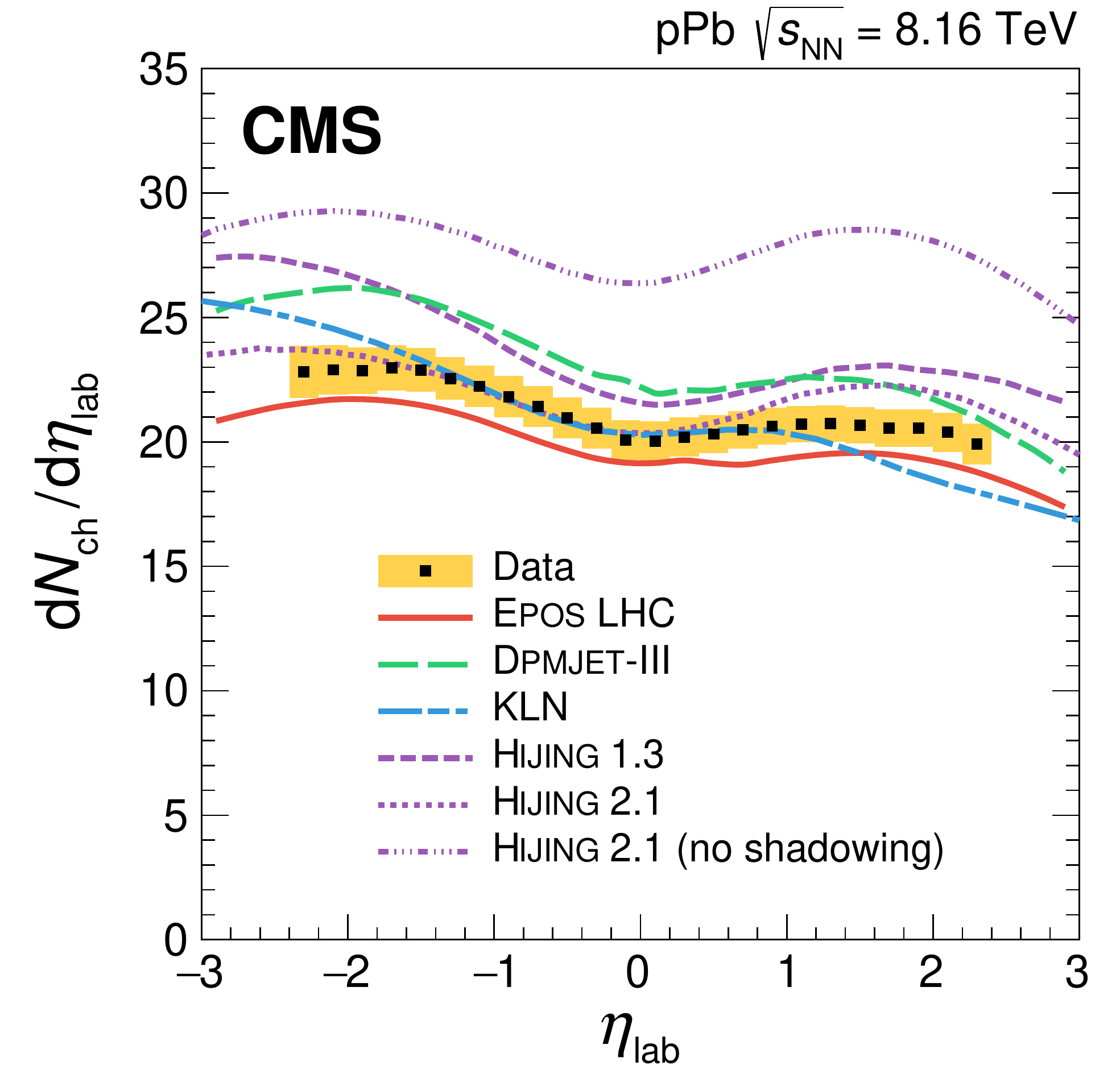}
\caption{Distributions of the pseudorapidity density of charged hadrons in the
region $\abs{\etalab} < 2.4$ for NSD \pPb collisions at 8.16\TeV (squares)
compared to predictions from the MC event generators \EPOS
LHC~\cite{Werner:2005jf,Pierog:2013ria} (v3400), \hijing~\cite{Wang:1991hta}
(versions 1.3~\cite{Gyulassy:1994ew} and 2.1~\cite{Xu:2012au}), and
\dpmjet~\cite{Roesler:2000he}, as well as from the KLN
model~\cite{Dumitru:2011wq}. The shaded boxes around the data points indicate
their systematic uncertainties. The proton beam goes in the positive \etalab
direction. }
\label{fig:8tev-predictions}
\end{figure}

Figure~\ref{fig:8tev-predictions} shows a comparison between the measurement at
8.16\TeV and theoretical calculations from the \hijing (versions 1.3 and 2.1),
\EPOS LHC (v3400), and \dpmjet MC generators, and the KLN model. The \hijing
and \EPOS generators were tuned to data from RHIC and the LHC, respectively.
Calculations from \hijing~2.1, a two-component model that combines perturbative
QCD descriptions of hard parton scatterings with a string excitation model for
soft interactions, agree with the experimental data in the region $-0.5
< {\etalab} < 1.5$ when the nuclear modification of the initial parton
distributions (shadowing) is included in the calculation. The \hijing~1.3
calculation overpredicts the particle density because it has an older
implementation of the gluon shadowing effects. The importance of shadowing can
be assessed using the comparison of \mbox{\hijing~2.1} simulations generated
with and without this physics process included. The results are significantly
higher than the data when shadowing is disabled. The KLN parton saturation
model combines Glauber modeling of the collision geometry with a simple model
for the unintegrated parton distributions that accounts for the existence of
a saturation momentum scale~\cite{Kharzeev:2001gp,Kharzeev:2001yq}. It
describes the particle density accurately for $\abs{\etalab} < 1$ but overall
shows a steeper increase of density versus \etalab than observed in the data,
similar to what was observed in the comparisons to the PHOBOS deuteron-gold
(dAu) data at 200\GeV~\cite{Back:2003hx} and ALICE data at
5.02\TeV~\cite{ALICE:2012xs}. The \dpmjet generator, commonly used in the
description of cosmic ray, nucleon-nucleon, and nucleon-nucleus interactions,
is based on the dual parton model~\cite{Capella:1992yb}, which generates soft
hadronic interactions by considering the expansion of nonperturbative QCD in
the limit where the number of color and flavor states are
large~\cite{Veneziano:1974ag}. This generator is found to predict both
a steeper increase versus \etalab and a higher particle density over the
measured \etalab interval. The \EPOS generator, which is based on the
Gribov--Regge theory and includes the effect of collective hadronization in
hadron-hadron scattering, was found to describe \pp data up to
13\TeV~\cite{Khachatryan:2015jna}, but underpredicts the observed \dnchdetalab
by a roughly constant factor over the entire measured range for \pPb at
8.16\TeV.

One of the main goals of the heavy ion studies is to understand hadron
production in the extremely dense medium formed in \NN collisions. One way to
approach this goal is to consider a direct comparison between the
charged-hadron multiplicity density in minimum bias \pp and \pA collisions, reference
systems for particle production in the absence of a QGP, and central \NN
collisions (the most extreme type of collisions with the highest particle
multiplicities). The comparison is made by dividing \dnchdetacm by the number
of participating nucleons, \npart, determined by a Glauber model
calculation~\cite{Miller:2007ri,Khachatryan:2016odn}. This normalization is the
one assumed in two-component models (\eg \ \hijing) for the bulk of the
particle production.

In order to compare particle production in \pPb collisions to that in symmetric
collision systems such as \pp or \NN, the rapidity shift due to the asymmetric
beam energies must be taken into account. The average charged-hadron
multiplicity density at midrapidity in the center-of-mass frame,
\avednchdetacm, in \pPb collisions is calculated by integrating the data in the
interval $-0.035 < \etalab < 0.965$, corresponding to $\abs{\etacm} < 0.5$ for
massless particles. A correction is applied to account for the massless
assumption entering the calculation of the pseudorapidity shift: 0.1 and 0.2\%
for the 5.02\TeV and 8.16\TeV analyses, respectively, as obtained from the
\EPOS generator. The 1\% variation in the results, obtained when this
correction is evaluated from \hijing, is quoted as an additional uncertainty
for the \avednchdetacm results. In the range $\abs{\etacm} < 0.5$, values of
$17.31 \pm 0.01\stat\pm 0.59\syst$ and $20.10 \pm
0.01\stat\pm 0.85\syst$ are obtained for \pPb
collisions at $\sqrtsNN = 5.02$ and 8.16\TeV, respectively.

\begin{figure}[t!bhp]
\centering
\includegraphics[width=0.60\textwidth]{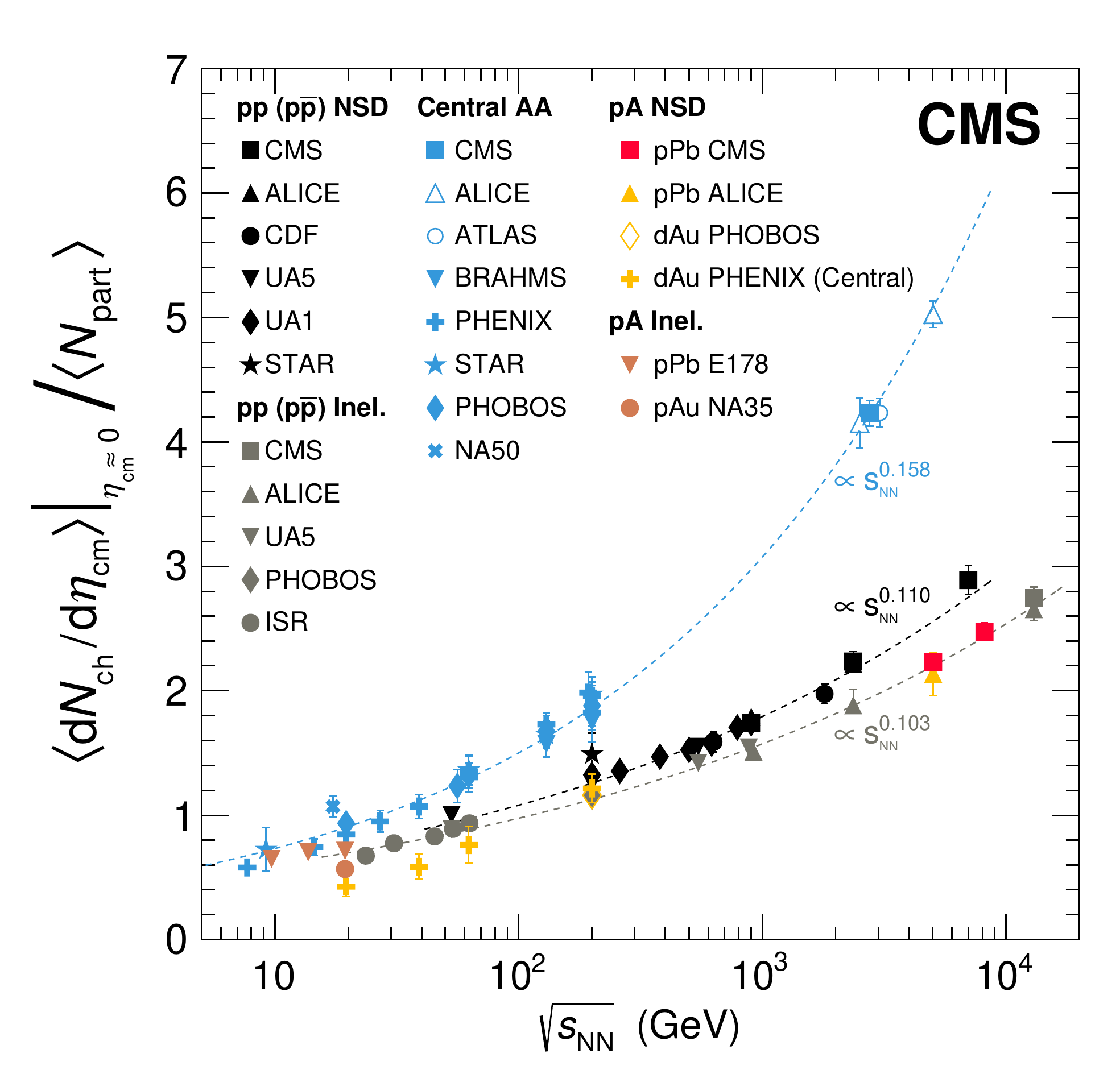}
\caption{Comparison of the measured $\dnchdetacm$ at midrapidity, scaled by the
number of participating nucleons (\npart) in
\pPb~\cite{ALICE:2012xs,Elias:1979cp},
pAu~\cite{Alber:1997sn},
dAu~\cite{Back:2003hx,Adare:2015bua,Aidala:2017pup} and
central heavy ion collisions~\cite{Alver:2010ck,Abreu:2002fw,Adler:2001yq,
Bearden:2001xw,Bearden:2001qq,Adcox:2000sp,Back:2000gw,Back:2001bq,
Back:2002wb,Aamodt:2010pb,ATLAS:2011ag,Chatrchyan:2011pb,Adare:2015bua,
Abelev:2009bw,Abelev:2008ab}, as well as
NSD~\cite{Albajar:1989an,Alpgard:1982kx,Alner:1986xu,Abelev:2008ab,
Abe:1989td,Khachatryan:2010xs,Khachatryan:2010us} and
inelastic~\cite{Thome:1977ky,Alner:1986xu,Aamodt:2010ft,Alver:2010ck,
Khachatryan:2015jna} \pp collisions. The AA data points at $\sqrtsNN = 2.76$\TeV
have been shifted horizontally for visibility. The dashed curves, included to
guide the eye, correspond to a fit to the data points using the same
functional form as in Refs.~\cite{Aamodt:2010ft,Aamodt:2010pb}. }
\label{fig:dndeta_vs_cme}
\end{figure}

Figure~\ref{fig:dndeta_vs_cme} shows the dependence of normalized \dnchdetacm
on the collision energy for various collision systems and event selections. The
NSD \pA results are found to be lower than those from central AA
collisions~\cite{Alver:2010ck,Abreu:2002fw,Adler:2001yq, Bearden:2001xw,
Bearden:2001qq,Adcox:2000sp,Back:2000gw,Back:2001bq, Back:2002wb,Aamodt:2010pb,
ATLAS:2011ag,Chatrchyan:2011pb,Adare:2015bua,Abelev:2009bw,Abelev:2008ab}
($s_{_{\mathrm{NN}}}^{0.158}$ dependence) and NSD \pp collisions
($s_{_{\mathrm{NN}}}^{0.110}$ dependence) at similar center-of-mass energies,
but coincide with the trend observed in inelastic \pp collisions
($s_{_{\mathrm{NN}}}^{0.103}$ dependence). While the difference between the NSD
\pp and \pA results could be attributed to  non-QGP nuclear effects, the
similarity between the NSD \pA and total inelastic \pp is yet to be understood.

\section{Summary}

The pseudorapidity distributions of primary charged hadrons have been measured
by the CMS experiment at the LHC in proton-lead collisions at $\sqrtsNN = 5.02$
and 8.16\TeV. Based on pairs of pixel clusters from two different layers of the
barrel region of the CMS pixel detector, the distributions have been obtained
for NSD \pPb events at both collision energies. The measured \dnchdetalab
distribution at 5.02\TeV is consistent with published results by the ALICE
Collaboration. At 8.16\TeV, the measured \dnchdetalab distribution is higher
than the predictions of \EPOS LHC, but significantly lower than the predictions
from the \hijing~1.3 and \dpmjet event generators. At $\etalab \approx 0$, the
measured distributions are in good agreement with calculations from the KLN
gluon saturation model and predictions from the \hijing~2.1 event generator
with the effects of gluon shadowing included. The charged-hadron multiplicity
densities in the nucleon-nucleon center-of-mass frame, \avednchdetacm, are
$17.31 \pm 0.01\stat\pm 0.59\syst$ and $20.10 \pm
0.01\stat\pm 0.85\syst$ at $\sqrtsNN = 5.02$ and
8.16\TeV, respectively. When comparing the average charged-particle density per
participant nucleon for \pp, \pA, and \NN collisions as a function of collision
energy, the \pA results are found to be below those in central \NN collisions
and NSD \pp collisions, but coincide with the trend seen in inelastic \pp
collisions. These results represent the first measurement of hadron production
at this new center-of-mass energy frontier in nuclear collisions, and provide
constraints for the understanding of nonperturbative QCD effects in high-energy
nuclear collisions.

\begin{acknowledgments}
We congratulate our colleagues in the CERN accelerator departments for the excellent performance of the LHC and thank the technical and administrative staffs at CERN and at other CMS institutes for their contributions to the success of the CMS effort. In addition, we gratefully acknowledge the computing centers and personnel of the Worldwide LHC Computing Grid for delivering so effectively the computing infrastructure essential to our analyses. Finally, we acknowledge the enduring support for the construction and operation of the LHC and the CMS detector provided by the following funding agencies: BMWFW and FWF (Austria); FNRS and FWO (Belgium); CNPq, CAPES, FAPERJ, and FAPESP (Brazil); MES (Bulgaria); CERN; CAS, MoST, and NSFC (China); COLCIENCIAS (Colombia); MSES and CSF (Croatia); RPF (Cyprus); SENESCYT (Ecuador); MoER, ERC IUT, and ERDF (Estonia); Academy of Finland, MEC, and HIP (Finland); CEA and CNRS/IN2P3 (France); BMBF, DFG, and HGF (Germany); GSRT (Greece); OTKA and NIH (Hungary); DAE and DST (India); IPM (Iran); SFI (Ireland); INFN (Italy); MSIP and NRF (Republic of Korea); LAS (Lithuania); MOE and UM (Malaysia); BUAP, CINVESTAV, CONACYT, LNS, SEP, and UASLP-FAI (Mexico); MBIE (New Zealand); PAEC (Pakistan); MSHE and NSC (Poland); FCT (Portugal); JINR (Dubna); MON, RosAtom, RAS, RFBR and RAEP (Russia); MESTD (Serbia); SEIDI, CPAN, PCTI and FEDER (Spain); Swiss Funding Agencies (Switzerland); MST (Taipei); ThEPCenter, IPST, STAR, and NSTDA (Thailand); TUBITAK and TAEK (Turkey); NASU and SFFR (Ukraine); STFC (United Kingdom); DOE and NSF (USA).

\hyphenation{Rachada-pisek} Individuals have received support from the Marie-Curie program and the European Research Council and Horizon 2020 Grant, contract No. 675440 (European Union); the Leventis Foundation; the A. P. Sloan Foundation; the Alexander von Humboldt Foundation; the Belgian Federal Science Policy Office; the Fonds pour la Formation \`a la Recherche dans l'Industrie et dans l'Agriculture (FRIA-Belgium); the Agentschap voor Innovatie door Wetenschap en Technologie (IWT-Belgium); the Ministry of Education, Youth and Sports (MEYS) of the Czech Republic; the Council of Science and Industrial Research, India; the HOMING PLUS program of the Foundation for Polish Science, cofinanced from European Union, Regional Development Fund, the Mobility Plus program of the Ministry of Science and Higher Education, the National Science Center (Poland), contracts Harmonia 2014/14/M/ST2/00428, Opus 2014/13/B/ST2/02543, 2014/15/B/ST2/03998, and 2015/19/B/ST2/02861, Sonata-bis 2012/07/E/ST2/01406; the National Priorities Research Program by Qatar National Research Fund; the Programa Severo Ochoa del Principado de Asturias; the Thalis and Aristeia programs cofinanced by EU-ESF and the Greek NSRF; the Rachadapisek Sompot Fund for Postdoctoral Fellowship, Chulalongkorn University and the Chulalongkorn Academic into Its 2nd Century Project Advancement Project (Thailand); the Welch Foundation, contract C-1845; and the Weston Havens Foundation (USA).
\end{acknowledgments}

\bibliography{auto_generated}
\cleardoublepage \appendix\section{The CMS Collaboration \label{app:collab}}\begin{sloppypar}\hyphenpenalty=5000\widowpenalty=500\clubpenalty=5000\textbf{Yerevan Physics Institute,  Yerevan,  Armenia}\\*[0pt]
A.M.~Sirunyan, A.~Tumasyan
\vskip\cmsinstskip
\textbf{Institut f\"{u}r Hochenergiephysik,  Wien,  Austria}\\*[0pt]
W.~Adam, F.~Ambrogi, E.~Asilar, T.~Bergauer, J.~Brandstetter, E.~Brondolin, M.~Dragicevic, J.~Er\"{o}, M.~Flechl, M.~Friedl, R.~Fr\"{u}hwirth\cmsAuthorMark{1}, V.M.~Ghete, J.~Grossmann, J.~Hrubec, M.~Jeitler\cmsAuthorMark{1}, A.~K\"{o}nig, N.~Krammer, I.~Kr\"{a}tschmer, D.~Liko, T.~Madlener, I.~Mikulec, E.~Pree, N.~Rad, H.~Rohringer, J.~Schieck\cmsAuthorMark{1}, R.~Sch\"{o}fbeck, M.~Spanring, D.~Spitzbart, W.~Waltenberger, J.~Wittmann, C.-E.~Wulz\cmsAuthorMark{1}, M.~Zarucki
% ...
\vskip\cmsinstskip
% ...
1:~~Also at Vienna University of Technology, Vienna, Austria\\
% ...

\end{sloppypar}
\end{document}